\documentclass[superscriptaddress,12pt]{revtex4}
\usepackage{amsfonts}
\usepackage{graphicx}
\usepackage{marvosym}
\begin{document}
\title{Shortcuts to adiabatic passage for multiqubit controlled phase gate {\footnote{Yan Liang $\cdot$ Xin Ji ({\large \Letter})
 Department of Physics, College of Science, Yanbian University,
\\ Yanji, Jilin 133002, People's Republic of China
 e-mail: jixin@ybu.edu.cn}}}
\author{\textbf{Yan Liang $\cdot$ Xin Ji} }

\begin{abstract}
\noindent \textbf{{Abstract}}: We propose an alternative scheme of
shortcuts to quantum phase gate in a much shorter time based on
the approach of ¡°Lewis-Riesenfeld invariants¡± in cavity quantum
electronic dynamics (QED) systems. This scheme can be used to
perform one-qubit phase gate, two-qubit controlled phase gate and
also multiqubit controlled phase gate. The strict numerical
simulation for some quantum gates are given, and demonstrate that
the total operation time of our scheme is shorter than previous
schemes and very robustness against decoherence.
\\ {\bf{Keywords:}} {Shortcuts to adiabatic passage } $\cdot$ {one-qubit phase gate} $\cdot$
{multiqubit controlled phase gate}
\end{abstract}
\maketitle
\section{Introduction}\label{section1}
 It is well known that quantum gates play a significant role in quantum
computing,~and any quantum gate operation can be decomposed into a
series of one-qubit gates and two-qubit conditional gates, such as
one-qubit phase gate, and two-qubit controlled-NOT gate
\cite{DPD1995,AB1995}. Recently, a number of schemes have been
proposed to perform quantum logic gates using optical devices
\cite{YFH2004}, QED system \cite{SBZ2013}, quantum
dot\cite{BQH2002}, ion trap and superconducting devices
\cite{JIC1995,M2001,CPY2003,CPY2006}. Moreover, the
implementations of two-qubit conditional gates in experiment have
been proposed \cite{MAN2000,SLB2001}. However, the controlled
phase gates with more than three qubits is difficult in
experimental implementation. Even though a $N$-qubit controlled
phase gate could be decomposed into one- and two-qubit gates, it
would be extremely complex for a practical problem, even worse, it
would increase the total operation time so that the decoherence
arising which will destroy the quantum system eventually.
Recently, a great many schemes have been proposed to perform
quantum logical gate via adiabatic passage
\cite{HGK2004,ZKF2002,BRS2013,DDB2014}. For example, Hayato Goto
et al. implemented the multiqubit controlled unitary gate by
adiabatic passage with an optical cavity\cite{HGK2004}. Zheng
implemented a $\pi$ phase gate through the adiabatic evolution
\cite{SBZ2005}. Rydberg-interaction gates with adiabatic passage
was proposed in \cite{DDB2014}. All these schemes are based on
adiabatic passage technique, because this method allows the
initial state evolve along the dark state to the target state
accurately.

 However, the adiabatic condition usually requires a
relatively long interaction time and then slows down the speed of
the system evolution, and finally the dissipation caused by
decoherence, noise, and losses would destroy the expected
dynamics. Therefore, accelerating the dynamics towards the target
outcome would be the most reasonable and effective way to actually
fight against the decoherence. Thus, the shortcuts to adiabatic
passage for various reliable, fast, and robust schemes have been
drawn a lot of attentions in both theory and experiment
\cite{ARXD2012,XASA2010,KPYR2011,AC2013,MYLJ2014,YYQJ2014,AFTS2012,JXPP2011}.
However the shortcuts to logical gates have not been fully
studied. Chen et al. \cite{YHC2014} proposed a scheme of shortcuts
to performing a $\pi$ phase gate through designing the particular
resonant laser pulses by the invariant-based inverse engineering.
It was the only scheme for quantum logic gates based on shortcuts
to adiabatic passage in cavity QED systems.

In this paper, we construct an effective shortcuts to adiabatic
passage to perform one-qubit phase gate, two-qubit controlled
phase gate, three-qubit controlled phase gate and also multiqubit
controlled phase gate based on the Lewis-Riesenfeld invariants and
quantum Zeno dynamics. The logical gates in our scheme can be
performed in a much shorter time than that based on adiabatic
passage technique. Moreover, this scheme is insensitive to the
decoherence caused by spontaneous emission and photon leakage
which is demonstrated by the strict numerical simulation.

~This paper is structured as follows: In Sec. II, we give a brief
description about Lewis-Riesenfeld invariants. In Sec. III, we
construct a shortcuts to one-qubit phase gate. Two-qubit
controlled phase gate, three-qubit controlled phase gate and
multiqubit controlled phase gate are presented in Sec. IV. In Sec.
V we give the numerical simulation and feasibility analysis for
our schemes. The conclusion appears in Sec. VI.

\begin{figure}[htb]\centering
\includegraphics[width=11cm]{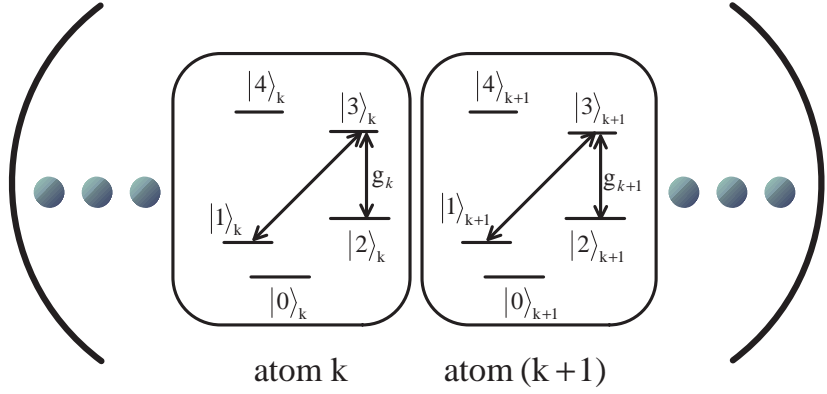}
\caption{The schematic setup for multiqubit phase gate. The $N$
five levels atoms which have the same level structure are trapped
in a single mode optical cavity. }
\end{figure}

\section{Lewis-Riesenfeld invariants}

We first give a brief description about Lewis-Riesenfeld
invariants theory \cite{HRL1969,MAL2009}. A quantum system is
governed by a time-dependent Hamiltonian $H(t)$, and the
corresponding time-dependent Hermitian invariant $I(t)$ satisfies
\begin{eqnarray}\label{1}
i\hbar \frac{\partial I(t)}{\partial t}&=&[H(t),I(t)].
\end{eqnarray}
The solution of the time-dependent Schr\"odinger equation $i\hbar$
$\frac{\partial |\Psi(t)\rangle}{\partial t}$ $
=H(t)|\Psi(t)\rangle$ can be expressed by a superposition of
invariant $I(t)$ dynamical modes $|\Phi_{n}(t)\rangle$:
\begin{eqnarray}\label{2}
|\Psi(t)\rangle&=&\sum_n C_n e^{i\alpha_n}|\Phi_{n}(t)\rangle,
\end{eqnarray}
where $C_n$ is a time-independent amplitude, $\alpha_n$ is the
Lewis-Riesenfeld phase, $|\Phi_{n}(t)\rangle$ are orthonormal
eigenvectors of the invariant $I(t)$, satisfying
$I(t)|\Phi_{n}(t)\rangle=\lambda_n|\Phi_{n}(t)\rangle$, with
$\lambda_n$ real constants. And the Lewis-Riesenfeld phases are
defined as
\begin{eqnarray}\label{3}
\alpha_n(t)&=&\frac{1}{\hbar}\int_0^\prime dt^\prime
\langle\Phi_n(t^\prime )|i\hbar\frac{\partial}{\partial t^\prime
}-H(t^\prime )|\Phi_n(t^\prime )\rangle.
\end{eqnarray}

\section{Shortcuts to adiabatic passage for one-qubit phase gate}

 The schematic setup for our scheme is
shown in Fig. 1, $N$ identical five-level atoms trapped in a
single mode optical cavity. Every atom possesses three ground
states $|0\rangle$, $|1\rangle$, $|2\rangle$ and two excited
states $|3\rangle$, $|4\rangle$. The state $|2\rangle$ and
$|3\rangle$ is strongly coupled with the cavity mode field, and
the other transitions $|1\rangle\rightarrow|4\rangle$,
$|2\rangle\rightarrow|4\rangle$, $|1\rangle\rightarrow|3\rangle$,
$|0\rangle\rightarrow|3\rangle$ are resonant with the classical
laser field.

 We now consider the one-qubit $\pi$ phase gate. In this case,
 only one qubit is trapped in the single mode optical cavity.
 Choosing the initial state of the qubit is
 $|\Psi_0\rangle=\alpha|0\rangle+\beta|1\rangle$, after performing
 the $\pi$ phase gate, the outcome state becomes:
\begin{eqnarray}\label{4}
|\Psi\rangle=\alpha|0\rangle-\beta|1\rangle.
\end{eqnarray}

In the following, we explain the detail of how to construct the
shortcuts to adiabatic passage for one-qubit $\pi$ phase gate. We
choose the laser pulses resonant with the transition
$|1\rangle\rightarrow|4\rangle$ and
$|2\rangle\rightarrow|4\rangle$ transitions, and the corresponding
Rabi frequencies are denoted by $\Omega_1(t)$ and $\Omega_2(t)$,
respectively. The interaction Hamiltonian in the interaction
picture is given by ($\hbar=1$)
\begin{eqnarray}\label{5}
H(t)=\Omega_1(t)|4\rangle\langle1|+\Omega_2(t)|4\rangle\langle2|+\rm
H.c.
\end{eqnarray}
To speed up the gate performing by the dynamics of invariant based
inverse engineering, we need to find out the Hermitian invariant
operator $I(t)$, which satisfies
 $i\hbar \frac{\partial I(t)}{\partial t}=[H(t),I(t)]$, and here $H(t)$ possesses SU(2)
dynamical symmetry, so $I(t)$ can be easily given by
\cite{XCE2011}
\begin{eqnarray}\label{6}
I(t)=\chi\left(\cos\gamma\sin\beta|4\rangle\langle1|+\cos\gamma\cos\beta|4\rangle\langle2|+i\sin\gamma|2\rangle\langle1|
\right),
\end{eqnarray}
$\chi$ is an arbitrary constant with units of frequency to keep
$I(t)$ with dimensions of energy, $\gamma$, and $\beta$ are
time-dependent auxilary parameters which satisfy the equations
\begin{eqnarray}\label{7}
\dot{\gamma}&=&\Omega_1(t)\cos\beta-\Omega_2(t)\sin\beta,\nonumber\\
\dot{\beta}&=&\tan\gamma(\Omega_2(t)\cos\beta+\Omega_1(t)\sin\beta).
\end{eqnarray}
From Eq. (7) we can derive the expressions of $\Omega_1(t)$ and
$\Omega_2(t)$ as follow:
\begin{eqnarray}\label{8}
\Omega_1(t)&=&\dot{\beta}\cot\gamma\sin\beta+\dot{\gamma}\cos\beta,\nonumber\\
\Omega_2(t)&=&\dot{\beta}\cot\gamma\cos\beta-\dot{\gamma}\sin\beta.
\end{eqnarray}
The eigenstates of the invariant $I(t)$ are
\begin{eqnarray}\label{9}
|\Phi_0(t)\rangle&=&\cos\gamma\cos\beta|1\rangle-i\sin\gamma|4\rangle-\cos\gamma\sin\beta|2\rangle,\nonumber\\
|\Phi_\pm(t)\rangle&=&\frac{1}{\sqrt{2}}[(\sin\gamma\cos\beta\pm
i\sin\beta)|1\rangle+i\cos\gamma|4\rangle-(\sin\gamma\sin\beta\mp
i\cos\beta)|2\rangle].
\end{eqnarray}
The solution of the Schr\"{o}dinger equation $i\hbar$
$\frac{\partial |\Psi(t)\rangle}{\partial t}$ $
=H(t)|\Psi(t)\rangle$ can be written with the eigenstates of
$I(t)$ as
\begin{eqnarray}\label{10}
|\Psi(t)\rangle&=&\sum_{n=0,\pm}C_ne^{i\alpha_n}|\Phi_n(t)\rangle,
\end{eqnarray}
where $\alpha_n(t)$ are the  Lewis-Riesenfeld phases presented in
Eq. (3), and in this case, $C_n=\langle\Phi_n(0)|1\rangle$.

In order to generate a $\pi$ phase on the state $|1\rangle$, we
choose the parameters as
\begin{eqnarray}\label{11}
\gamma(t)&=&\epsilon,~~~~~~~~~\beta(t)=\pi t/t_f,
\end{eqnarray}
with $\epsilon$ is a time-independent small value and $t_f$ is the
total operation time. Then, we obtain
\begin{eqnarray}\label{12}
\Omega_1(t)&=&\frac{\pi}{t_f}\cot\epsilon\sin\frac{\pi t}{t_f},\nonumber\\
\Omega_2(t)&=&\frac{\pi}{t_f}\cot\epsilon\cos\frac{\pi t}{t_f}.
\end{eqnarray}
When $t=t_f$,
\begin{eqnarray}\label{13}
|\Psi(t_f)\rangle&=&(-\cos^2\epsilon-\sin^2\epsilon\cos\alpha)|1\rangle+(-i\sin\epsilon\cos\epsilon+i\sin\epsilon\cos\epsilon\cos\alpha)|4\rangle\cr
&&-\sin\epsilon\sin\alpha|2\rangle,
\end{eqnarray}
where $\alpha=\pi/\sin\epsilon=|\alpha_\pm|$. When we choose
$\alpha=2N\pi(N=1,2,3...)$, $|\Psi(t_f)\rangle=-|1\rangle$. On the
other hand, the state $|0\rangle$ does not participate in the
evolvtion. Therefore, the $\pi$ phase gate can be achieved
\begin{eqnarray}\label{14}
|\Psi_0\rangle=\alpha|0\rangle+\beta|1\rangle\rightarrow|\Psi\rangle=\alpha|0\rangle-\beta|1\rangle.
\end{eqnarray}

\section{Shortcuts to adiabatic passage for controlled phase gate}

\subsection{Two-qubit controlled $\pi$ phase gate}

In this section, a two-qubit controlled $\pi$ phase gate is
proposed. We consider two identical five-level atoms trapped in a
single mode optical cavity as shown in Fig. 1. The initial state
of two atoms is defined as
\begin{eqnarray}\label{15}
|\Psi_0\rangle=\alpha_{00}|00\rangle+\alpha_{01}|01\rangle+\alpha_{10}|10\rangle+\alpha_{11}|11\rangle,
\end{eqnarray}
where $\alpha_{ij}(i,j=0,1)$ denote the probability amplitude of
the state $|ij\rangle$. After performing the controlled $\pi$
phase gate, the output is
\begin{eqnarray}\label{16}
|\Psi\rangle=\alpha_{00}|00\rangle+\alpha_{01}|01\rangle+\alpha_{10}|10\rangle-\alpha_{11}|11\rangle,
\end{eqnarray}
where the first atom is control qubit, and the second atom acts as
target qubit. In order to construct the shortcuts to adiabatic
passage for two-qubit controlled $\pi$ phase gate, there are
mainly three steps.

Step 1: The second atom state $|1\rangle_2$ is transferred to
$-|2\rangle_2$ by the shortcuts with laser pulses resonant with
$|1\rangle_2\rightarrow|4\rangle_2$ and
$|2\rangle_2\rightarrow|4\rangle_2$, and the corresponding Rabi
frequencies are denoted by $\Omega_1(t)$ and $\Omega_2(t)$. Using
the similar method mentioned in Sec. III, and the only difference
is that we choose $\gamma(t)=\epsilon$ and $\beta(t)=\frac{\pi
t}{2t_f}$, then obtain
\begin{eqnarray}\label{17}
\Omega_1(t)&=&\frac{\pi}{2t_f}\cot\epsilon\sin\frac{\pi t}{2t_f},\nonumber\\
\Omega_2(t)&=&\frac{\pi}{2t_f}\cot\epsilon\cos\frac{\pi t}{2t_f}.
\end{eqnarray}
When $t=t_f$ and $\alpha=2N\pi(N=1,2,3...)$ , the transition
$|1\rangle_2\rightarrow-|2\rangle_2$ can be achieved. Then the
initial state $|\Psi_0\rangle$ becomes
\begin{eqnarray}\label{18}
|\Psi_1\rangle=\alpha_{00}|00\rangle-\alpha_{01}|02\rangle+\alpha_{10}|10\rangle-\alpha_{11}|12\rangle.
\end{eqnarray}

Step 2: The state $|12\rangle$ generates a $\pi$ phase by the
shortcuts and then becomes $-|12\rangle$ with laser pulses
resonant with the first atom $|1\rangle_1\rightarrow|3\rangle_1$
and the second atom $|1\rangle_2\rightarrow|3\rangle_2$, and the
corresponding Rabi frequencies denoted by $\Omega^{(1)}(t)$ and
$\Omega^{(2)}(t)$. By means of shortcuts, the state
$|\Psi_1\rangle$ becomes
\begin{eqnarray}\label{19}
|\Psi_2\rangle=\alpha_{00}|00\rangle-\alpha_{01}|02\rangle+\alpha_{10}|10\rangle+\alpha_{11}|12\rangle.
\end{eqnarray}
The detail of this step will be explained later.

 Step 3: The same with step 1, the second atom state
$|2\rangle_2$ is transferred back to $-|1\rangle_2$ by the
shortcuts with laser pulses resonant with
$|2\rangle_2\rightarrow|4\rangle_2$ and
$|1\rangle_2\rightarrow|4\rangle_2$, and the corresponding Rabi
frequencies are denoted by
$\Omega_2(t)=\frac{\pi}{2t_f}\cot\epsilon\sin\frac{\pi t}{2t_f}
$and $\Omega_1(t)=\frac{\pi}{2t_f}\cot\epsilon\cos\frac{\pi
t}{2t_f}$, when $t=t_f$ and $\alpha=2N\pi(N=1,2,3...)$, we can
obtain
\begin{eqnarray}\label{20}
|\Psi_3\rangle=\alpha_{00}|00\rangle+\alpha_{01}|01\rangle+\alpha_{10}|10\rangle-\alpha_{11}|11\rangle.
\end{eqnarray}
Thus, the two-qubit controlled $\pi$ phase gate can be achieved.

In the following, we explain how to realize step 2 in detail. We
have choose the laser pulses resonant with the first atom
$|1\rangle_1\rightarrow|3\rangle_1$ transition and the second atom
$|1\rangle_2\rightarrow|3\rangle_2$ transition with the
corresponding Rabi frequencies denoted by $\Omega^{(1)}(t)$ and
$\Omega^{(2)}(t)$. The Hamiltonian for this step is given by
\begin{eqnarray}\label{21}
H(t)=\Omega^{(1)}(t)|3\rangle_1\langle1|+\Omega^{(2)}(t)|3\rangle_2\langle1|+g_1a_1|3\rangle_1\langle2|+g_2a_2|3\rangle_2\langle2|+
\rm H.c,
\end{eqnarray}
where $g_{1,2}$ are the coupling constants between atoms and
cavity field modes, and $a_{1,2}$ are the annihilation operators
of photons. We choose $g_1=g_2$ and $a_1=a_2$ for simplicity. In
this case, if the system is in $|12\rangle_{1,2}|0\rangle_c$, the
evolution subspace can be spanned by
\begin{eqnarray}\label{22}
|\varphi_1\rangle&=&|12\rangle_{1,2}|0\rangle_c,~~~ |\varphi_2\rangle=|32\rangle_{1,2}|0\rangle_c,~~~ |\varphi_3\rangle=|22\rangle_{1,2}|1\rangle_c,\nonumber\\
|\varphi_4\rangle&=&|23\rangle_{1,2}|0\rangle_c,~~~
|\varphi_5\rangle=|21\rangle_{1,2}|0\rangle_c,
\end{eqnarray}
where $|0\rangle_c$ and $|1\rangle_c$ denote the photon number
state in the cavity field. With the help of quantum Zeno dynamics,
the effective Hamiltonian is given by \cite{YYQJ2014}
\begin{eqnarray}\label{23}
H_{eff}(t)=\frac{1}{\sqrt{2}}|\mu\rangle(\Omega^{(1)}(t)\langle\varphi_1|+\Omega^{(2)}(t)\langle\varphi_5|+
\rm{H.c)},
\end{eqnarray}
where
$|\mu\rangle=\frac{1}{\sqrt{2}}(-|\varphi_2\rangle+|\varphi_4\rangle)$.
It is obvious that the effective Hamiltonian $H_{eff}(t)$
possesses the SU(2) dynamical symmetriy, too. Thus, we can use the
same method presented in Sec. III. Choosing $\gamma=\epsilon$ and
$t_f=\frac{\pi t}{t_f}$, thus,
$\Omega^{(1)}(t)=\frac{\pi}{t_f}\cot\epsilon\sin\frac{\pi t}{t_f}$
and $\Omega^{(2)}(t)=\frac{\pi}{t_f}\cot\epsilon\cos\frac{\pi
t}{t_f}$. When $t=t_f$, $|12\rangle\rightarrow-|12\rangle$ can be
achieved.

If the initial state of the system is
$|10\rangle_{1,2}|0\rangle_c$, the vectors of the system evolution
subspace are given by
\begin{eqnarray}\label{24}
|\phi_1\rangle&=&|10\rangle_{1,2}|0\rangle_c,~~~
|\phi_2\rangle=|30\rangle_{1,2}|0\rangle_c,~~~
|\phi_3\rangle=|20\rangle_{1,2}|1\rangle_c.
\end{eqnarray}
For this case, the initial state $|10\rangle_{1,2}|0\rangle_c$
evolves along the dark state
\begin{eqnarray}\label{25}
|\phi_{dark}\rangle=g|\phi_1\rangle-\Omega^{(1)}(t)|\phi_3\rangle.
\end{eqnarray}
With the parameters above, when $t=t_f$, $\Omega^{(1)}(t)=0$, and
then $|\phi_{dark}\rangle=|\phi_1\rangle$. It is obvious that the
state $|00\rangle$ and $|02\rangle$ do not change any more in this
step. Thus the state in Eq. (19) can be obtained.

\subsection{ Three-qubit controlled $\pi$ phase gate}

The three-qubit controlled $\pi$ phase gate can be described as
follow: The three atoms initial state is given by
\begin{eqnarray}\label{26}
|\Psi_{0}\rangle=\sum_{l_1,l_2,l_3=0,1}\alpha_{l_1l_2l_3}|l_1l_2l_3\rangle,
\end{eqnarray}
 where $\alpha_{l_1l_2l_3}$ denote the probability amplitude of
the three five-level atoms state $|l_1l_2l_3\rangle
(l_1,l_2,l_3=0,1)$. After performing the controlled $\pi$ phase
gate, the output becomes
\begin{eqnarray}\label{27}
|\Psi\rangle=\sum_{l_1,l_2,l_3=0,1}e^{il_1l_2l_3\pi}\alpha_{l_1l_2l_3}|l_1l_2l_3\rangle,
\end{eqnarray}
with atom 1 and atom 2 are the two control qubits and atom 3 is
the target qubit. In order to construct the shortcuts to adiabatic
passage for three-qubit controlled $\pi$ phase gate, there are
mainly four steps.

Step 1: The third atom state $|1\rangle_3$ is transferred to
$-|2\rangle_3$ by the shortcuts with laser pulses resonant with
$|1\rangle_3\rightarrow|4\rangle_3$ and
$|2\rangle_3\rightarrow|4\rangle_3$ transitions, and the
corresponding Rabi frequencies are denoted by $\Omega_1(t)$ and
$\Omega_2(t)$. Similar to the method mentioned in the step 1 of
two-qubit controlled $\pi$ phase gate, we also choose
$\gamma(t)=\epsilon$ and $\beta(t)=\frac{\pi t}{2t_f}$, when
$t=t_f$ we can obtain
\begin{eqnarray}\label{28}
|\Psi_{1}\rangle=\sum_{l_1,l_2=0,1}(\alpha_{l_1l_20}|l_1l_20\rangle-\alpha_{l_1l_21}|l_1l_22\rangle).
\end{eqnarray}
Step 2: The state $|l_112\rangle$ is transferred to
$-|l_121\rangle$ with laser pulses resonant with the second atom
$|1\rangle_2\rightarrow|3\rangle_2$ transition and the third atom
$|1\rangle_3\rightarrow|3\rangle_3$ transition with the
corresponding Rabi frequencies are denoted by $\Omega^{(2)}(t)$
and $\Omega^{(3)}(t)$. Then, the state $|\Psi_1\rangle$ becomes
\begin{eqnarray}\label{29}
|\Psi_{2}\rangle=\sum_{l_1,l_2=0,1}(\alpha_{l_1l_20}|l_1l_20\rangle-\alpha_{l_101}|l_102\rangle+\alpha_{l_111}|l_121\rangle).
\end{eqnarray}
The detail of this step will be explained later.

Step 3: The state $|021\rangle$ transferred to $-|021\rangle$ with
laser pulses resonant with the first atom
$|0\rangle_1\rightarrow|3\rangle_1$ transition and the second atom
$|0\rangle_2\rightarrow|3\rangle_2$ transition, and the
corresponding Rabi frequencies denoted by $\Omega'^{(1)}(t)$ and
$\Omega'^{(2)}(t)$. Then, the state $|\Psi_2\rangle$ becomes
\begin{eqnarray}\label{30}
|\Psi_{3}\rangle=\sum_{l_1,l_2=0,1}\alpha_{l_1l_20}|l_1l_20\rangle-\sum_{l_1=0,1}\alpha_{l_101}|l_102\rangle-\alpha_{011}|021\rangle+\alpha_{111}|121\rangle.
\end{eqnarray}
The detail of this step will be explained later, too.

Step 4: $|2\rangle_{2(3)}$ is back to $-|1\rangle_{2(3)}$ by the
shortcuts with the similar method to step 1. As a result, the
output is
\begin{eqnarray}\label{31}
|\Psi_{4}\rangle&=&\sum_{l_1,l_2=0,1}\alpha_{l_1l_20}|l_1l_20\rangle+\sum_{l_1=0,1}\alpha_{l_101}|l_101\rangle+\alpha_{011}|011\rangle-\alpha_{111}|111\rangle\cr
&&=\sum_{l_1,l_2,l_3=0,1}e^{il_1l_2l_3\pi}\alpha_{l_1l_2l_3}|l_1l_2l_3\rangle.
\end{eqnarray}
Thus, the three-qubit controlled $\pi$ phase gate can be realized.

In the following, we explain the step 2 in detail. We have choose
the laser pulses resonant with the second atom
$|1\rangle_2\rightarrow|3\rangle_2$ transition and the third atom
$|1\rangle_3\rightarrow|3\rangle_3$ transition, and the
corresponding Rabi frequencies denoted by $\Omega^{(2)}(t)$ and
$\Omega^{(3)}(t)$. The Hamiltonian is given by
\begin{eqnarray}\label{32}
H(t)=\Omega^{(2)}(t)|3\rangle_2\langle1|+\Omega^{(3)}(t)|3\rangle_3\langle1|+g_2a_2|3\rangle_2\langle2|+g_3a_3|3\rangle_3\langle2|+
\rm H.c,
\end{eqnarray}
where $g_{2,3}$ are the coupling constants between atoms and
cavity field modes, and $a_{2,3}$ are the annihilation operators
of photons. We choose $g_2=g_3$ and $a_2=a_3$ for simplicity. In
this case, if the second and the third atoms state is
$|12\rangle_{2,3}$ , the evolution subspace spanned by
\begin{eqnarray}\label{33}
|\varphi_1\rangle&=&|12\rangle_{2,3}|0\rangle_c,~~~ |\varphi_2\rangle=|32\rangle_{2,3}|0\rangle_c,~~~ |\varphi_3\rangle=|22\rangle_{2,3}|1\rangle_c,\nonumber\\
|\varphi_4\rangle&=&|23\rangle_{2,3}|0\rangle_c,~~~
|\varphi_5\rangle=|21\rangle_{2,3}|0\rangle_c.
\end{eqnarray}
where $|0\rangle_c$ and $|1\rangle_c$ denote the photon number
state in the cavity field. The same with step 2 in the section of
two-qubit controlled $\pi$ phase gate, we now choose the parameter
$\gamma(t)=\epsilon$, $\beta(t)=\frac{\pi t}{2tf}$ and
$\alpha=2N\pi$, when $t=t_f$, we can obtain $-|\varphi_5\rangle$,
and this step is successful. On the other hand, the other states
will not change in this step.

And then, we explain the step 3 in detail. In this step, we choose
the laser pulses resonant with the first atom
$|0\rangle_1\rightarrow|3\rangle_1$ transition and the second atom
$|0\rangle_2\rightarrow|3\rangle_2$ transition, and the
corresponding Rabi frequencies are denoted by $\Omega'^{(1)}(t)$
and $\Omega'^{(2)}(t)$. The Hamiltonian is given by
\begin{eqnarray}\label{34}
H(t)=\Omega'^{(1)}(t)|3\rangle_1\langle0|+\Omega'^{(2)}(t)|3\rangle_2\langle0|+g'_{1}a'_{1}|3\rangle_1\langle2|+g'_{2}a'_{2}|3\rangle_3\langle2|+
H.c,
\end{eqnarray}
where $g_{1',2'}$ are the coupling constants between atoms and
cavity field modes, and $a_{1',2'}$ are the annihilation operators
of photons. We choose $g_{1'}=g_{2'}$ and $a_{1'}=a_{2'}$ for
simplicity. The vectors in this step are
\begin{eqnarray}\label{35}
|\xi_1\rangle&=&|02\rangle_{1,2}|0\rangle_c,~~~ |\xi_2\rangle=|32\rangle_{1,2}|0\rangle_c,~~~ |\xi_3\rangle=|22\rangle_{1,2}|1\rangle_c,\nonumber\\
|\xi_4\rangle&=&|23\rangle_{1,2}|0\rangle_c,~~~
|\xi_5\rangle=|20\rangle_{1,2}|0\rangle_c.
\end{eqnarray}
The same with the method described in step 2 in the section of
two-qubit controlled $\pi$ phase gate, we now choose the parameter
$\gamma(t)=\epsilon$, $\beta(t)=\frac{\pi t}{tf}$ and
$\alpha=2N\pi$, when $t=t_f$, we can obtain $-|02\rangle_{1,2}$,
and the other states will not change in this step.

\subsection{ Multiqubit controlled $\pi$ phase gate}

We consider $n+1$ atoms are trapped in a single mode cavity as
shown in Fig. 1. In general, $(n+1)$-qubit controlled $\pi$ phase
gate can be described as follow: The $(n+1)$ atoms is in the
initial state
\begin{eqnarray}\label{36}
|\Psi_{0}\rangle=\sum_{l_1,l_2...l_{n+1}=0,1}\alpha_{l_1l_2...l_{n+1}}|l_1l_2...l_{n+1}\rangle.
\end{eqnarray}
After performing the $(n+1)$-qubit controlled $\pi$ phase gate, we
can obtain
\begin{eqnarray}\label{37}
|\Psi\rangle=\sum_{l_1,l_2...l_{n+1}=0,1}e^{i\pi
l_1l_2...l_{n+1}}\alpha_{l_1l_2...l_{n+1}}|l_1l_2...l_{n+1}\rangle.
\end{eqnarray}
Here, the first $n$ qubits are the control qubits, and the last
qubit is target qubit. In order to construct the shortcuts to
$(n+1)$-qubit controlled $\pi$ phase gate, there are four steps.

 Step 1: $|1\rangle_{n+1}$ is transferred to $-|2\rangle_{n+1}$ by
 the laser pulses resonant with the $(n+1)th$ atom $|1\rangle_{n+1}\rightarrow|4\rangle_{n+1}$ and
 $|2\rangle_{n+1}\rightarrow|4\rangle_{n+1}$ transitions. Then the initial
 state becomes
\begin{eqnarray}\label{38}
|\Psi_{1}\rangle=\sum_{l_1,l_2...l_n=0,1}(\alpha_{l_1l_2...l_n0}|l_1l_2...l_n0\rangle-\alpha_{l_1l_2...l_n1}|l_1l_2...l_n2\rangle).
\end{eqnarray}
Step 2: The state $|l_1l_2...l_{n-1}12\rangle$ transferred to
$-|l_1l_2...l_{n-1}21\rangle$ with laser pulses resonant with the
$nth$ atom $|1\rangle_n\rightarrow|3\rangle_n$ transition and the
$(n+1)th$ atom $|1\rangle_{n+1}\rightarrow|3\rangle_{n+1}$
transition, and the corresponding Rabi frequencies denoted by
$\Omega^{(n)}(t)$ and $\Omega^{(n+1)}(t)$. Then, the state
$|\Psi_1\rangle$ becomes
\begin{eqnarray}\label{39}
|\Psi_{2}\rangle&=&\sum_{l_1,l_2...l_n=0,1}\alpha_{l_1l_2...l_n0}|l_1l_2...l_n0\rangle-\sum_{l_1,l_2...l_{n-1}=0,1}\alpha_{l_1l_2...l_{n-1}01}|l_1l_2...l_{n-1}02\rangle\cr
&&+\sum_{l_1,l_2...l_{n-1}=0,1}\alpha_{l_1l_2...l_{n-1}11}|l_1l_2...l_{n-1}21\rangle.
\end{eqnarray}

Step 3: The state $|l_1l_2...l_{n-1}21\rangle$ transferred to
$-|l_1l_2...l_{n-1}21\rangle$, here $l_1,l_2...l_{n-1}$ satisfied
the condition $l_1\cdot l_2\cdot...\cdot l_{n-1}=0$. Choosing the
laser pulses resonant with the $kth$ atom
$|0\rangle_k\rightarrow|3\rangle_k$ $(k=1,2...n-1)$ transition and
the $nth$ atom $|0\rangle_n\rightarrow|3\rangle_n$ transition, and
the corresponding Rabi frequencies denoted by $\Omega'^{(k)}(t)$
and $\Omega'^{(n)}(t)$. Then, the state $|\Psi_2\rangle$ becomes
\begin{eqnarray}\label{40}
|\Psi_{3}\rangle&=&\sum_{l_1,l_2...l_n=0,1}\alpha_{l_1l_2...l_n0}|l_1l_2...l_n0\rangle-\sum_{l_1,l_2...l_{n-1}=0,1}\alpha_{l_1l_2...l_{n-1}01}|l_1l_2...l_{n-1}02\rangle\cr
&&
-\sum_{l_1,l_2...l_{n-1}=0,1}\alpha_{l_1l_2...l_{n-1}11}|l_1l_2...l_{n-1}21\rangle_{(l_1\cdot
l_2\cdot...l_{n-1}=0)}+\alpha_{11...1}|1...21\rangle.
\end{eqnarray}
Step 4: $|2\rangle_{n(n+1)} $ is back to $-|1\rangle_{n(n+1)}$ by
the shortcuts. As a result, the output is
\begin{eqnarray}\label{41}
|\Psi_{4}\rangle&=&\sum_{l_1,l_2...l_n=0,1}\alpha_{l_1l_2...l_n0}|l_1l_2...l_n0\rangle+\sum_{l_1,l_2...l_{n-1}=0,1}\alpha_{l_1l_2...l_{n-1}01}|l_1l_2...l_{n-1}01\rangle\cr
&&
+\sum_{l_1,l_2...l_{n-1}=0,1}\alpha_{l_1l_2...l_{n-1}11}|l_1l_2...l_{n-1}11\rangle_{(l_1\cdot
l_2\cdot...l_{n-1}=0)}-\alpha_{11...1}|1...11\rangle\cr
&&=\sum_{l_1,l_2...l_{n+1}=0,1}e^{i\pi
l_1l_2...l_{n+1}}\alpha_{l_1l_2...l_{n+1}}|l_1l_2...l_{n+1}\rangle.
\end{eqnarray}
That is a $(n+1)$-qubit controlled $\pi$ phase gate.

\section{Numerical simulations and feasibility analysis}
In this section, we make the numerical simulations for one-qubit
$\pi$ phase gate and two-qubit controlled $\pi$ phase gate by
numerically solving the Schr$\ddot{\rm o}$dinger equations. We
also discuss the influence of spontaneous emission and decay of
cavity on fidelity.

\subsection{ One-qubit $\pi$ phase gate}

\begin{figure}
\scalebox{0.9}
{\includegraphics[width=3.0in,height=2.5in]{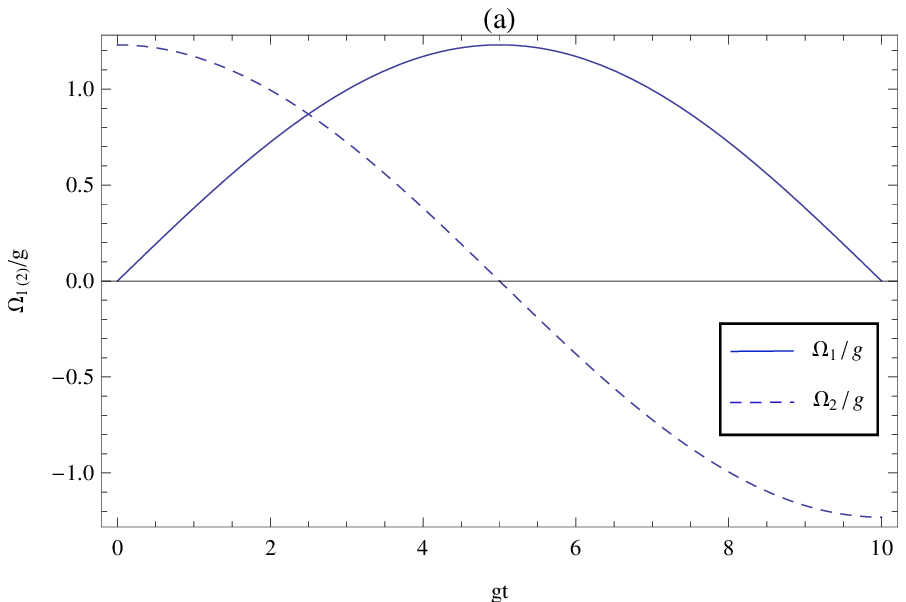}
\includegraphics[width=3.0in,height=2.5in]{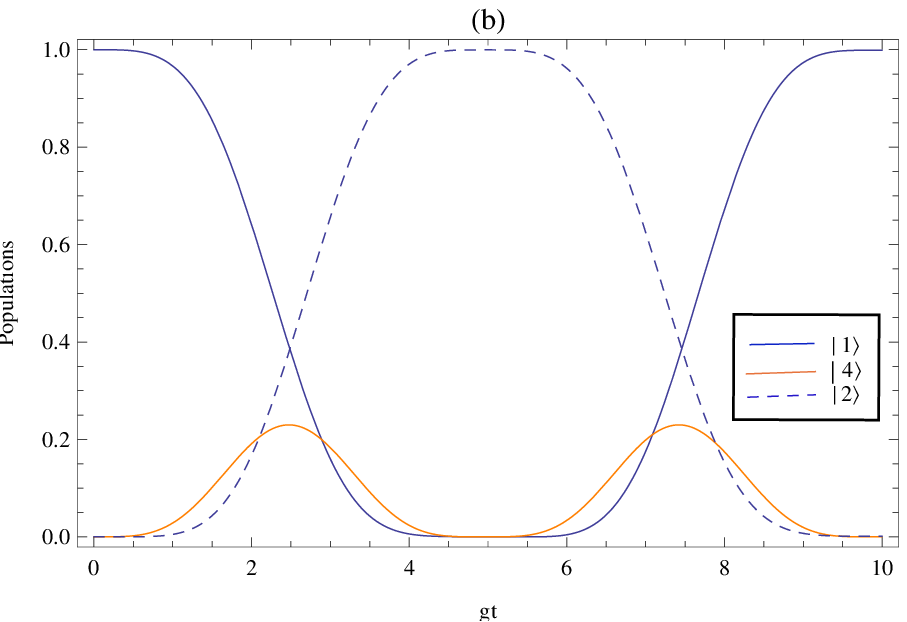}} \caption{(a)
Time dependence of $\Omega_{1(2)}(t)/g$ of the laser fields for
performing one-qubit $\pi$ phase gate. (b) Time evolutions of the
populations of corresponding system states. Here, the system
parameters are set to be $\epsilon= 0.25$ and $
t_{f}=10/g.$}\label{fig02}
\end{figure}

\begin{figure}
\scalebox{0.9}
{\includegraphics[width=3.0in,height=2.5in]{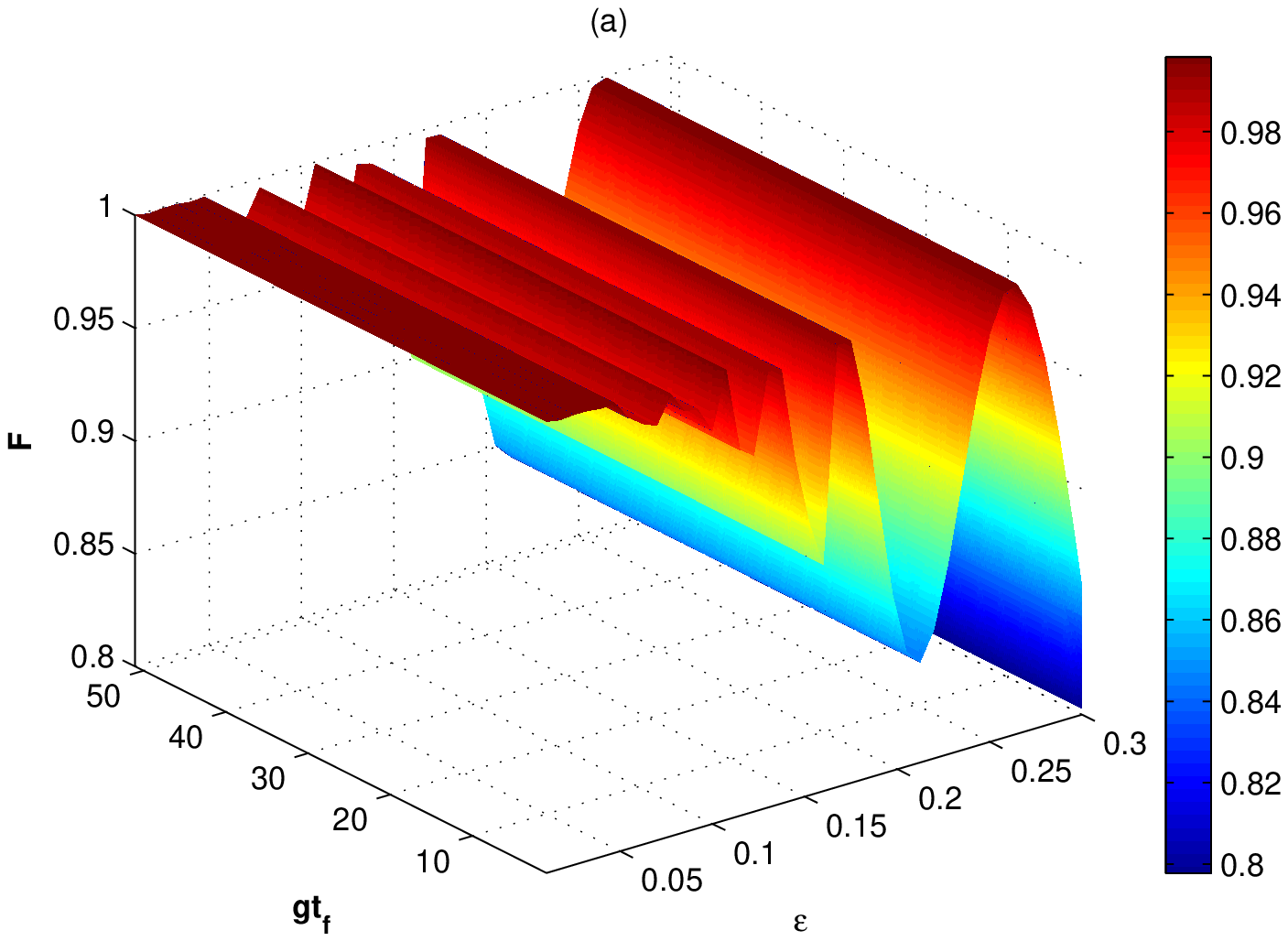}
\includegraphics[width=3.0in,height=2.5in]{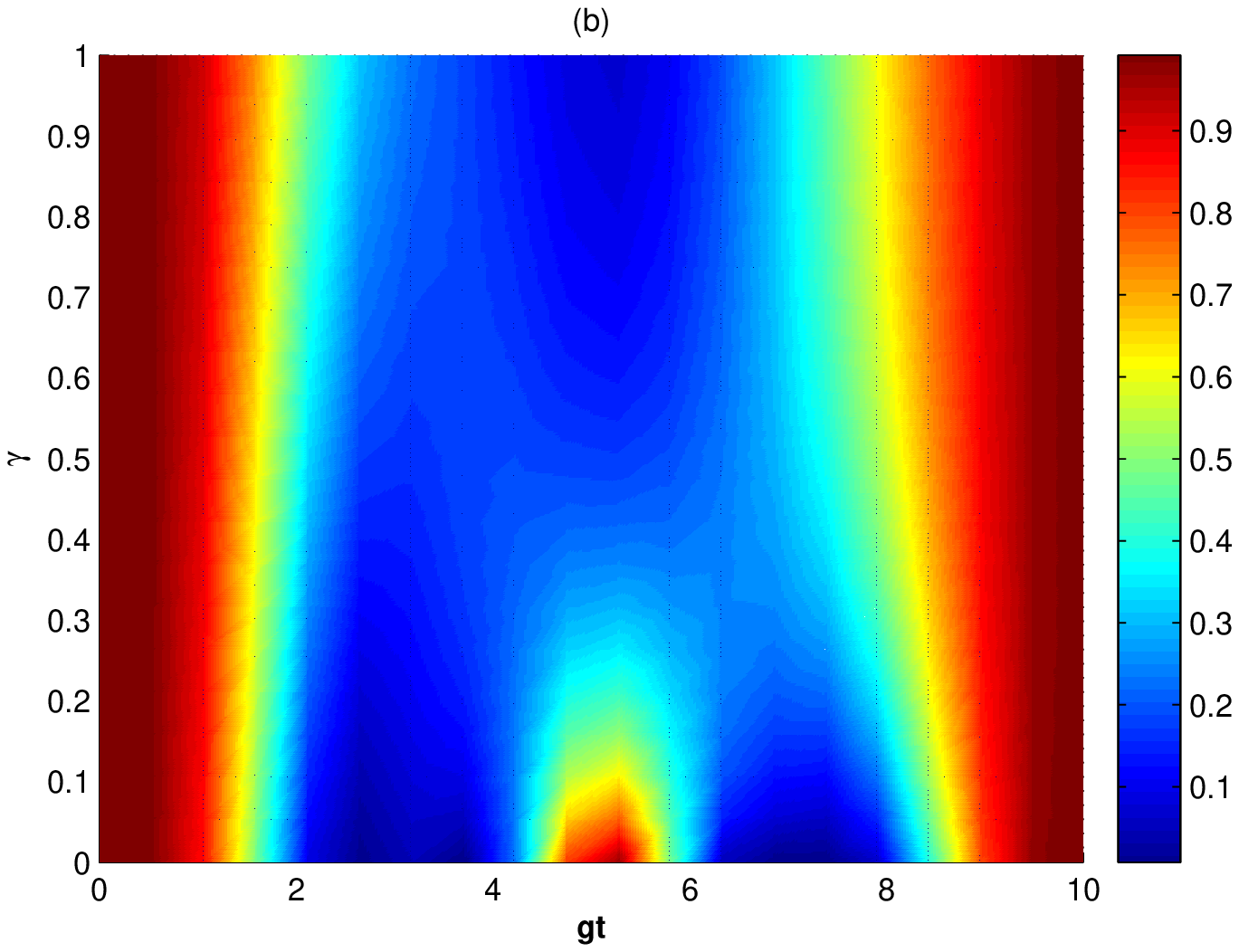}} \caption{(a)
The fidelity of one-qubit $\pi$ phase gate versus $\epsilon$ and
$t_f$ regardless of the atom decay. (b) The fidelity of one-qubit
$\pi$ phase gate versus the evolution time $t$ and the spontaneous
emission rate $\gamma$ of atom. The system parameters are set to
be $\epsilon= 0.25$ and $ t_{f}=10/g.$}\label{fig03}
\end{figure}

We consider the initial state of the single atom is given by:
\begin{eqnarray}\label{42}
|\Psi_0\rangle =\frac{1}{\sqrt{2}}(|0\rangle+|1\rangle).
\end{eqnarray}
Fig.~2(a) shows the scaled Rabi frequencies $\Omega_{1}(t)/g$ and
$\Omega_{2}(t)/g$ versus $gt$ when $\epsilon=0.25$ and
$gt_{f}=10$, where $\Omega_{1}(t)$ and $\Omega_{2}(t)$ is defined
in Eq. (12). The population curves of $ |1\rangle $, $|4\rangle$
and $|2\rangle$ versus $gt$ are depicted in Fig.~2(b). From
Fig.~2(b) we can see a perfect population transfer from the
initial state $|1\rangle$ and then back to $|1\rangle$ after the
whole involution, and generate a $\pi$ phase which can be known
from Eq. (13).~Through the above processes, we construct the
shortcuts for one-qubit $\pi$ phase gate successfully. As
expected, the final state is
\begin{eqnarray}\label{43}
|\Psi\rangle=\frac{1}{\sqrt{2}}(|0\rangle-|1\rangle).
\end{eqnarray}
Fig. 3(a) shows the fidelity $F(t_{f})= \langle-1|\Psi(t_{f})$ as
a function of $\epsilon$ and $gt_{f}$ when the initial state is
$|1\rangle$. Fig. 3(a) demonstrates that, the effect of $t_f$ on
fidelity can be ignored. In our scheme, we choose $\epsilon=0.25$,
and the fidelity can be higher than $99\%$.

Next, we investigate the influence of spontaneous emission of atom
on the gate fidelity. The evolution of the system is governed by
the master equation
\begin{eqnarray}\label{44}
\dot{\rho}&=&i[\rho,H]+\sum_{i=1,2}\frac{\gamma}{2}[|i\rangle\langle4|\rho|4\rangle\langle
i|-\frac{1}{2}(|4\rangle\langle4|\rho+\rho|4\rangle\langle4|)],
\end{eqnarray}
 $\gamma$ is the spontaneous emission rate of atom. We plot the fidelity $F(t)=
|\langle-1|\rho(t)|-1\rangle|$ as a function of the operation time
$t$ and spontaneous emission rate $\gamma$ in Fig.~3(b), with
$\rho(t)$ being density matrix at $t$ and $|-1\rangle$ being the
target state, and the other parameters are $\epsilon=0.25$ and
pulse duration $t_f=10/g$. The evolutions are governed by the
Hamiltonian defined in Eq. (5). From Fig. 3(b) we can see that,
when the total evolution time $t=t_f$, the fidelity of our scheme
can be higher than $99.7\%$, in other words, our scheme is
insensitive to the spontaneous emission of atom.

\subsection{ two-qubit $\pi$ phase gate}

\begin{figure}
\centering
\begin{minipage}{8cm}\includegraphics[width=4.0in,height=1.5in]{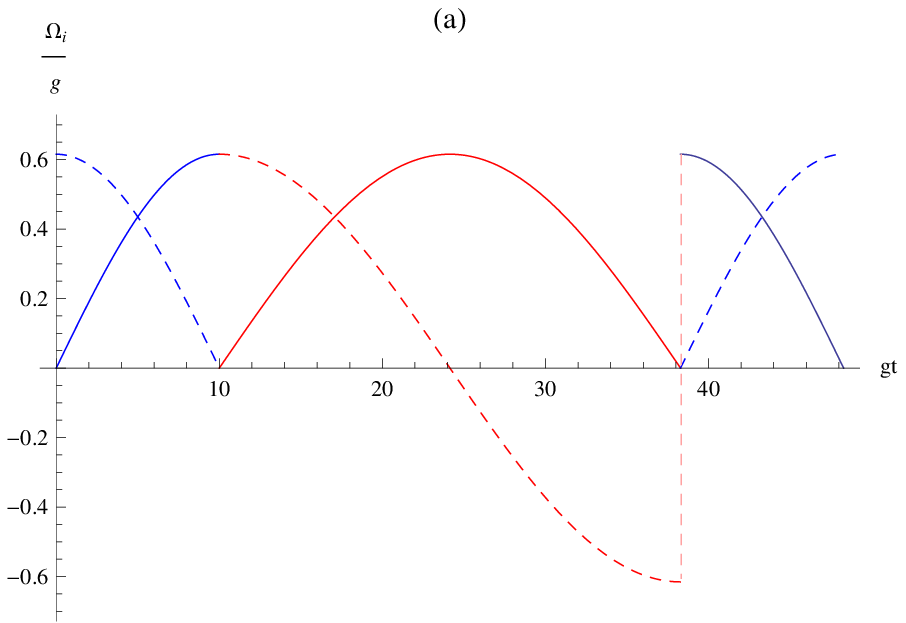}\end{minipage}

\begin{minipage}{8cm}\includegraphics[width=4.0in,height=1.5in]{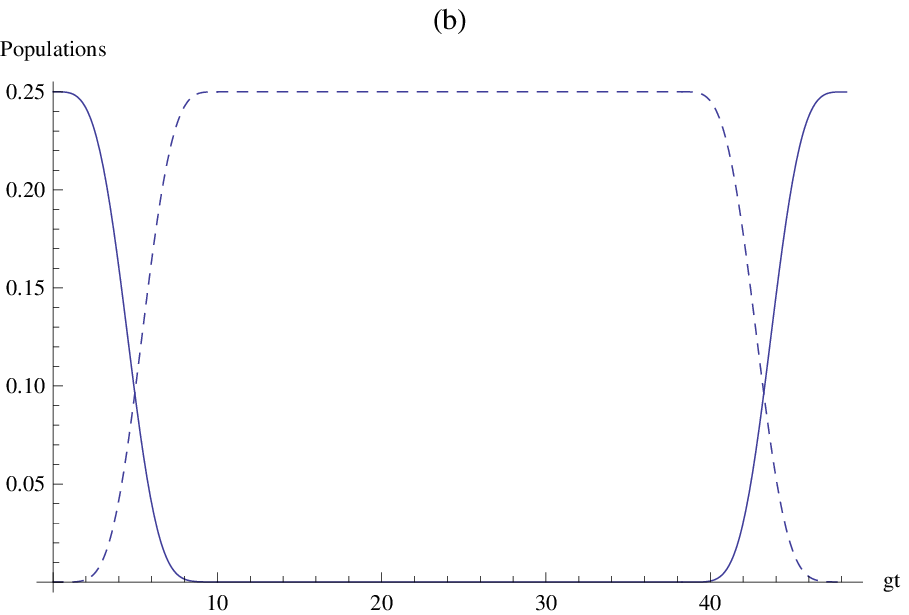}\end{minipage}

\begin{minipage}{8cm}\includegraphics[width=4.0in,height=1.5in]{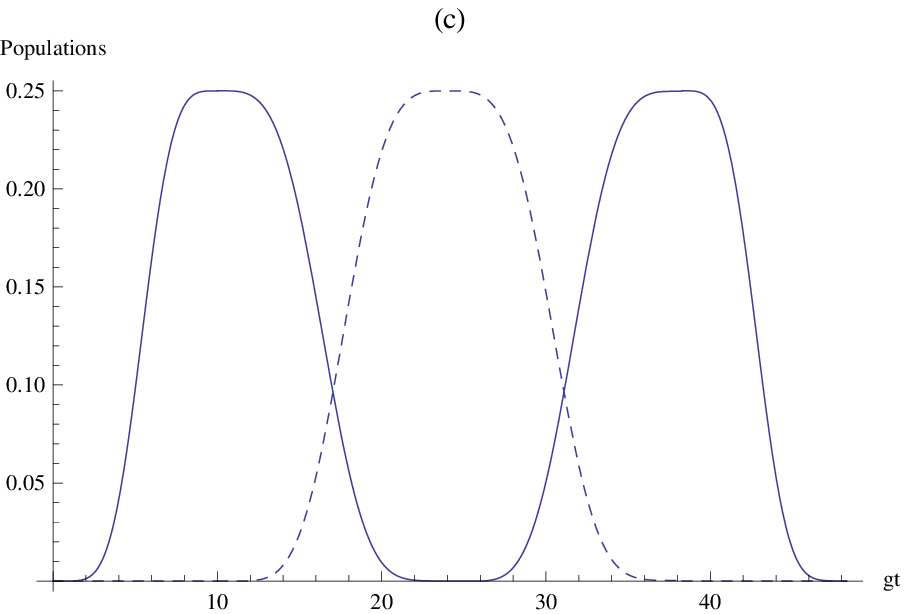}\end{minipage}

\caption{(a) Time dependence of $\Omega_{i}(t)/g$ of the laser
fields for two-qubit controlled $\pi$ phase gate with
$\Omega_i(t)=\Omega_1(t)$ (solid blue line), $\Omega_2(t)$ (dash
blue line), $\Omega^{(1)}(t)$ (solid red line), $\Omega^{(2)}(t)$
(dash red line). (b) Time evolutions of the populations of
corresponding system states $|01\rangle$ (solid blue line) and
$-|02\rangle$ (dash blue line). (c) Time evolutions of the
populations of corresponding system states $|12\rangle$ (solid
blue line) and $|21\rangle$ (dash blue line). The system
parameters are set to be $\epsilon= 0.25$ and $
t_{f}=10/g.$}\label{fig04}
\end{figure}

We consider the initial state of the two atom is given by:
\begin{eqnarray}\label{45}
|\Psi_0\rangle=\frac{1}{2}(|00\rangle+|01\rangle+|10\rangle+|11\rangle).
\end{eqnarray}

The coupling rate of atom and cavity field mode are chosen as
$g_1=g_2=g$. We depict the scaled Rabi frequencies
$\Omega_{1}(t)/g$, $\Omega_{2}(t)/g$ ,$\Omega^{(1)}(t)/g$ and
$\Omega^{(2)}(t)/g$ versus $gt$ in Fig. 4(a), and the other
parameters are chosen as $\epsilon=0.25$ and $gt_{f}=10$, where
$\Omega_{1}(t)$ and $\Omega_{2}(t)$ is defined in Eq. (17),
$\Omega^{(1)}(t)$ and $\Omega^{(2)}(t)$ are the same form with Eq.
(12). The population curves of $|01\rangle$ (solid blue line) and
$|02\rangle$ (dash blue line) versus $gt$  are shown in Fig. 4(b).
Fig. 4(c) shows the population of $|12\rangle$ (solid blue line)
and $|21\rangle$ (dash blue line) versus $gt$. ~Through the above
processes, we construct the shortcuts to two-qubit $\pi$ phase
gate successfully. As expected, the final state is
\begin{eqnarray}\label{46}
|\Psi\rangle=\frac{1}{2}(|00\rangle+|01\rangle+|10\rangle-|11\rangle).
\end{eqnarray}

We note that, in the step 2 of two-qubit controlled $\pi$ phase
gate, the vectors including
$|\varphi_3\rangle=|22\rangle_{2,3}|1\rangle_c$ in Eq. (33), i.e.
there is a photon in the cavity. Therefore, we must both
investigate the influence of cavity decay and spontaneous emission
on the gate fidelity. The evolution of the system is governed by
the master equation
\begin{eqnarray}\label{47}
\dot{\rho}&=&i[\rho,H]+\sum_{k=1}^5[L_k\rho
L_k^+-\frac{1}{2}(L_k^+L_k\rho+\rho L_k^+L_k)] ,
\end{eqnarray}
where $L_k$ and $L_k^+$ are the Lindblad operators \cite{MJK2011},
and they have the following form
\begin{eqnarray}\label{48}
L_1&=&\sqrt{\kappa}a,\nonumber~~~~~~~~~~~~~~~
L_2=\sqrt{\gamma_1}|1\rangle_2\langle3|,\nonumber~~~~~~~
L_3=\sqrt{\gamma_2}|1\rangle_3\langle3|,\nonumber\\
L_4&=&\sqrt{\gamma_3}|2\rangle_2\langle3|,~~~~~~~
L_3=\sqrt{\gamma_4}|2\rangle_3\langle3|,
\end{eqnarray}

where $\kappa$ is the decay rate of cavity and $\gamma_i
(i=1,2,3,4)$ are the corresponding spontaneous emission rates of
atoms, and $H$ is defined by Eq. (32). We choose
$\gamma_i=\gamma$. We plot the fidelity $F(t)=
|\langle-12|\rho(t)|-12\rangle|$ as a function of the operation
time $t$ and cavity decay rate $\kappa$ in Fig.~5(a)), and as a
function of the operation time $t$ and spontaneous emission rate
$\gamma$ in Fig.~5(b)), with $\rho(t)$ being density matrix at $t$
and $|-12\rangle$ being the target state, and the other parameters
are $\epsilon=0.25$ and pulse duration $t_f=20\sqrt{2}/g$. From
Fig. 5(a) and Fig. (b) we can see that, when the total evolution
time $t=t_f$, the fidelity of our scheme is closed to 1.
Therefore, our scheme is robust against the cavity decay and
spontaneous emission, and must be feasible in experiment.
\begin{figure}
\scalebox{0.9}
{\includegraphics[width=3.0in,height=2.5in]{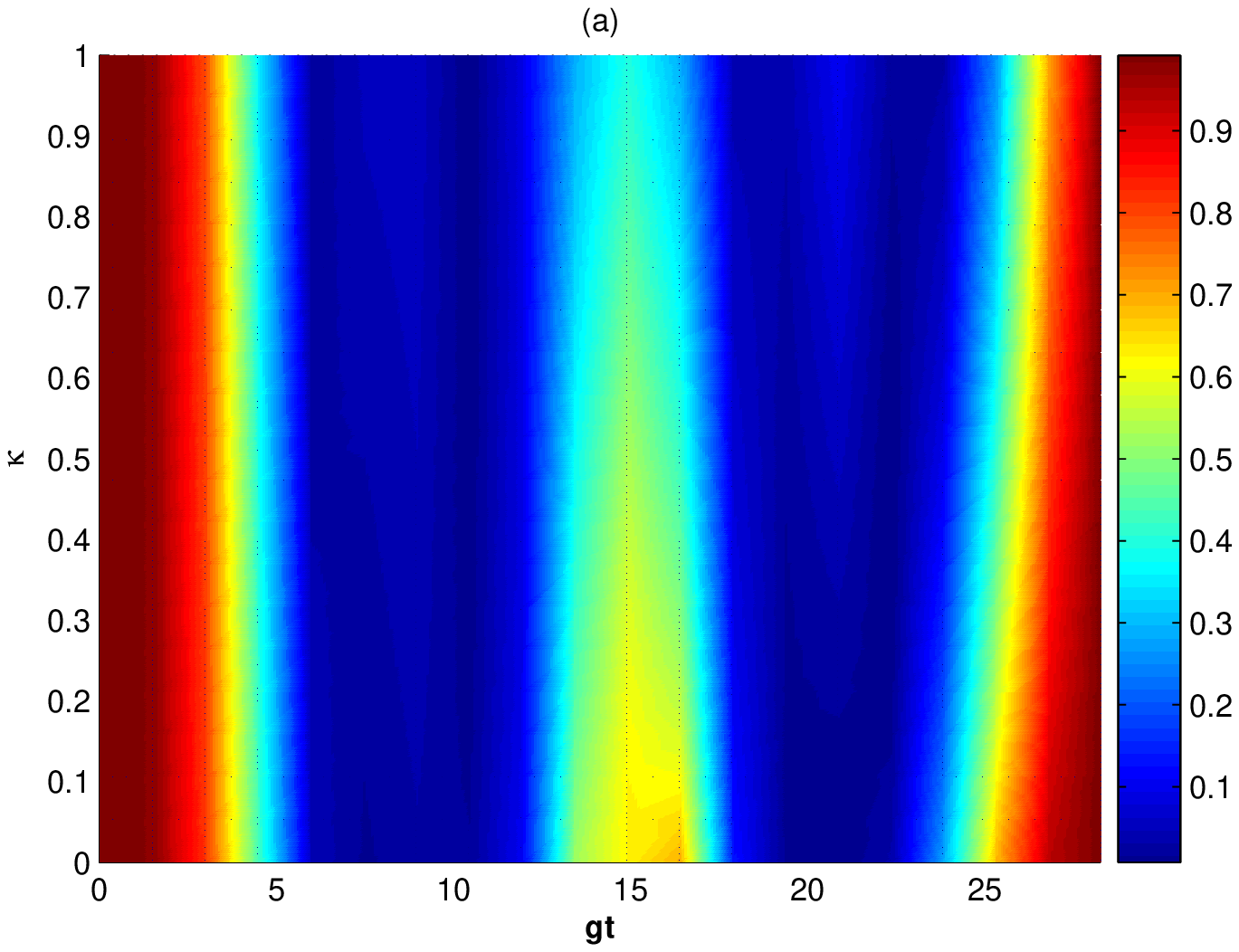}
\includegraphics[width=3.0in,height=2.5in]{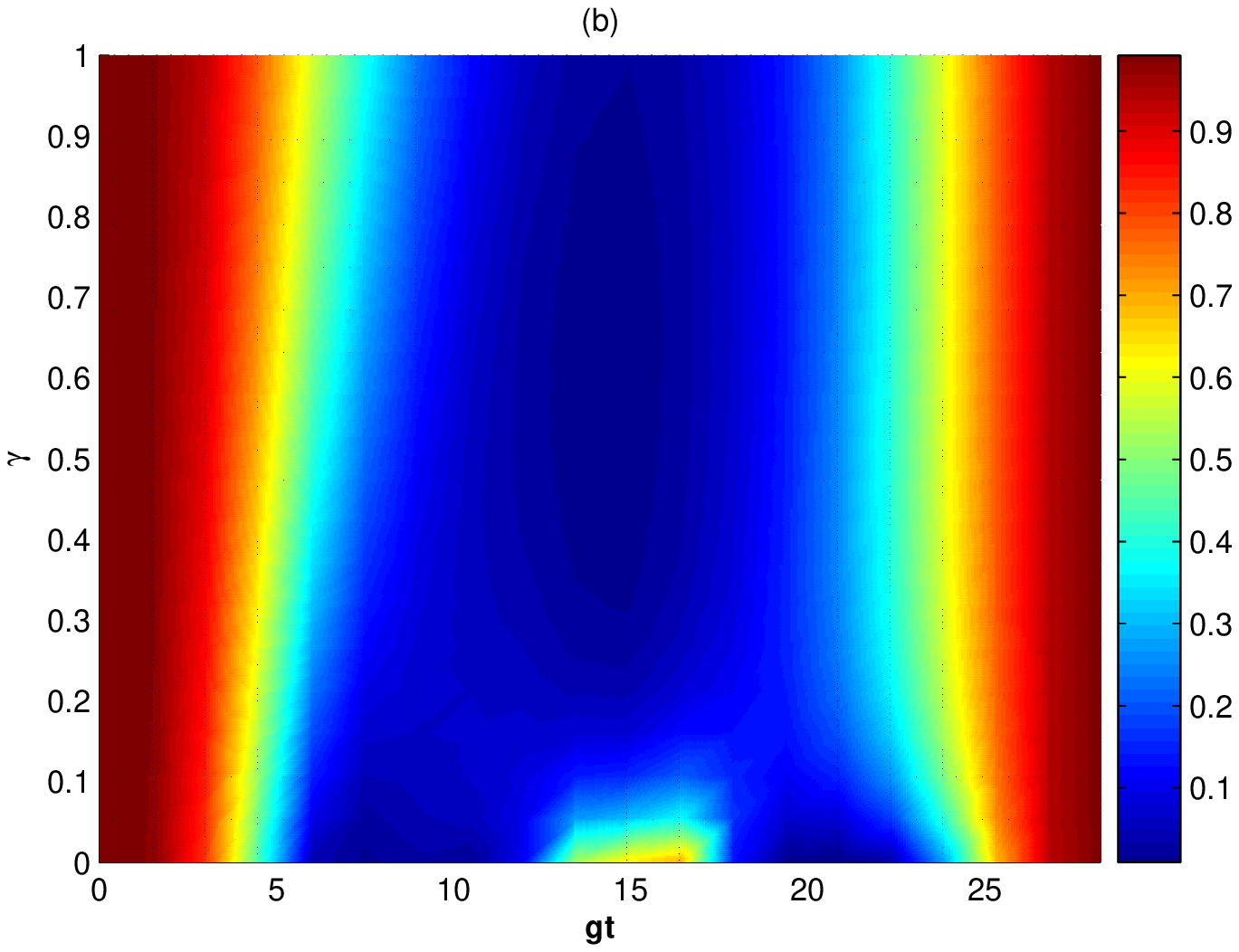}} \caption{(a)
The fidelity of the step 2 of two-qubit controlled $\pi$ phase
gate versus $\kappa$ and the evolution time $t $. (b) The fidelity
of the step 2 of two-qubit controlled $\pi$ phase gate versus the
spontaneous emission of atom $\gamma$ and the evolution time $t$.
The system parameters are set to be $\epsilon= 0.25$ and $
t_{f}=20\sqrt{2}/g.$}\label{fig02}
\end{figure}

We now analyze the feasibility in experiment for this scheme. The
appropriate atomic level configuration can be realized with
trapped ions and cavity QED systems \cite{SSD2000,JPH2002,LYX2003}
or with impurity levels in a solid, such as Pr$^{3+}$ ions in
Y$_2$SiO$_5$ crystal \cite{KI2001}, or nitrogen-vacancy color
center in diamond \cite{MSS2002}. In experiments, the cavity QED
parameters $(g,\kappa,\gamma)/2\pi=(750,3.5,2.62)$ MHz is
predicted to be available in an optical cavity \cite{SMS2005}. In
our scheme, when the cavity decay rate and the spontaneous
emission rate is comparable to atom cavity coupling constant $g$,
the fidelity is also higher than $99\%$. Thus, our scheme is
robust against both the cavity decay and atomic spontaneous
radiation and may be very promising within current experiment
technology.

\section{ Conlusion}

In summary, we have proposed a promising scheme to construct
shortcuts to perform one-qubit phase gate and muliqubit controlled
phase gate by invariant-based inverse engineering. Compared with
the previous work, the interaction time required for the gate
operation is much shorter than that with the method of adiabatic
passage. The shortcuts to our scheme is not only fast, but also
robust against the decoherence caused by atomic spontaneous
emission and cavity decay, so it can be a more reliable choice in
experiment.
\\

\begin{center}$\mathbf{ACKNOWLEDGMENTS}$\end{center}
This work was supported by the National Natural Science Foundation
of China under Grant Nos. 11464046 and 61465013.


\begin{thebibliography}{999}
\bibitem{DPD1995} DiVincenzo, D. P.: Two-bit gates are universal for quantum computation. Phys. Rev. A \textbf{51}, 1015 (1995).
\bibitem{AB1995} Barenco, A., Bennett, C. H., Cleve, R., DiVincenzo, D. P., Margolus, N., Shor, P., Sleator, T., Smolin,
J.,
 Weinfurter, H.: Elementary gates for quantum computation. Phys. Rev. A \textbf{52}, 3457 (1995).
\bibitem{YFH2004} Huang, Y. F., Ren, X. F., Zhang, Y. S., Duan, L. M., Guo, G. C.: Experimental teleportation of a quantum controlled-NOT gate. Phys. Rev. Lett. \textbf{93}, 240501 (2004).
\bibitem{SBZ2013} Zheng, S. B.: Implementation of toffoli gates with a single asymmetric Heisenberg XY interaction. Phys. Rev. A \textbf{87}, 042318 (2013).
\bibitem{BQH2002} Qiao, B., Ruda, H. E., Wang, J.: Multiqubit computing and error-avoiding codes in subspace
using quantum dots. J. Appl. Phys. \textbf{91}, 2524 (2002).
\bibitem{JIC1995}  Cirac, J. I., Zoller, P.: Quantum Computations with Cold Trapped Ions. Phys. Rev. Lett. \textbf{74}, 4091 (1995).
\bibitem{M2001} $\breve{\textrm{S}}$a$\breve{\textrm{s}}$ura, M., Bu$\breve{\textrm{z}}$ek, V.: Multiparticle entanglement
with quantum logic networks: Application to cold trapped ions.
Phys. Rev. A \textbf{64}, 012305 (2001).
\bibitem{CPY2003} Yang, C. P., Chun, S.: Possible realization of entanglement, logical gates, and quantum-information transfer
with superconducting-quantum-interference-device qubits in cavity
QED. Phys. Rev. A \textbf{67}, 042311 (2003).
\bibitem{CPY2006} Yang, C. P., Han, S.: Realization of an n-qubit controlled-U gate with superconducting quantum interference devices
or atoms in cavity QED. Phys. Rev. A \textbf{73}, 032317 (2006).
\bibitem{MAN2000}  Nielsen, M. A., Chuang, I. L.:  Quantum Computation and
Quantum Information. (Cambridge University Press, Cambridge,
2000).
\bibitem{SLB2001} DiVincenzo, D. P., Braunstein, S. L., Lo, H. K.: Scalable Quantum Computers (Wiley-
VCH, Berlin, 2001).
\bibitem{HGK2004} Goto, H., Ichimura, K.: Multiqubit controlled unitary gate by adiabatic passage with an optical cavity. Phys. Rev. A \textbf{70}, 012305 (2004).
\bibitem{ZKF2002} Kis, Z., Renzoni, F.: Qubit rotation by stimulated Raman adiabatic passage. Phys. Rev. A \textbf{65}, 032318 (2002).
\bibitem{BRS2013} Roussraux, B., Gu\'{e}rin, S., Vitanov, N. V.: Arbitrary qudit gates by adiabatic passage. Phys. Rev. A \textbf{87}, 032328(2013).
\bibitem{DDB2014} Rao, D. D. B., M$\o$lmer, K.: Robust Rydberg-interaction gates with adiabatic passage. Phys. Rev. A \textbf{89}, 030301(R) (2014).
\bibitem{SBZ2005} Zheng, S. B.: Nongeometric conditional phase shift via adiabatic evolution of dark eigenstates:
a new approach to quantum computation. Phys. Rev. Lett.
\textbf{95}, 080502 (2005).
\bibitem{ARXD2012} Ruschhaupt, A., Chen, X., Alonso, D., Muga,  J. G.: Optimally robust shortcuts to population inversion in two-level quantum systems. New J. Phys. \textbf{14}, 093040
(2012).
\bibitem{XASA2010} Chen, X., Lizuain, I., Ruschhaupt, A.,
Gu\'{e}ry-Odelin, D., Muga, J. G.: shortcut to adiabatic passage
in two- and three-level atoms. Phys. Rev. Lett. \textbf{105},
123003 (2010).
\bibitem{KPYR2011} Hoffmann, K. H., Salamon, P., Rezek, Y., Kosloff, R.: Time-optimal controls for frictionless cooling in harmonic traps. Euro.
Phys. Lett. \textbf{96}, 60015 (2011).
\bibitem{AC2013} del Campo, A.: Shortcuts to adiabaticity by counter-diabatic driving. Phys. Rev. Lett. \textbf{111},
100502 (2013).
\bibitem{MYLJ2014} Lu, M., Xia, Y., Shen, L. T., Song, J., An, N. B.: Shortcuts to adiabatic passage for population transfer and maximum entanglement creation
between two atoms in a cavity. Phys. Rev. A \textbf{89}, 012326
(2014).
\bibitem{YYQJ2014} Chen, Y. H., Xia, Y., Chen, Q. Q., Song, J.: Efficient shortcuts to adiabatic passage for fast population transfer in multiparticle systems. Phys. Rev. A \textbf{89}, 033856 (2014).
\bibitem{AFTS2012} Walther, A., Ziesel, F., Ruster, T., Dawkins, S. T., Ott, K.,
Hettrich, M., Singer, K., Schmidt-Kaler, F., Poschinger, U.:
Controlling fast transport of cold trapped ions. Phys. Rev. Lett.
\textbf{109}, 080501 (2012).
\bibitem{JXPP2011} Schaff, J. F., Song, X. L., Capuzzi, P., Vignolo, P., Labeyrie,
G.: Shortcut to adiabaticity for an interacting Bose-Einstein
condensate. Euro. Phys. Lett. \textbf{93}, 23001 (2011).
\bibitem{YHC2014} Cheng, Y. H., Xia, Y., Chen, Q. Q., Song, J.: Fast and noise-resistant implementation of quantum phase gates
and creation of quantum entangled states. arXiv preprint arXiv:
\textbf{1410}, 8285 (2014).
\bibitem{HRL1969} Lewis, H. R., Riesenfeld, W. B.: An exact quantum theory of the timeDependent harmonic oscillator and
of a charged particle in a timeDependent electromagnetic field. J.
Math. Phys. \textbf{10}, 1458 (1969).
\bibitem{MAL2009} Lohe, M. A.: Exact time dependence of solutions to the
time-dependent Schr¡§odinger equation. J. Phys. A: Math. and
Theor. \textbf{42}, 035307 (2009).
\bibitem{XCE2011} Chen, X., Torrontegui, E., Muga, J. G.: Lewis-Riesenfeld invariants and transitionless quantum driving. Phys. Rev. A \textbf{83}, 062116 (2011).
\bibitem{MJK2011} Kastoryano, M. J., Reiter, F., S\o rensen, A. S.: Dissipative preparation of entanglement in optical cavities. Phys. Rev. Lett. \textbf{106}, 090502 (2011).
\bibitem{SSD2000} Schneider, S., James, D., Milburn, G. J.: Quantum computation with hot trapped ions. J. Mod. Opt. \textbf{47}, 499 (2000).
\bibitem{JPH2002} Pachos, J., Walther, H.: Quantum computation with trapped ions in an optical cavity. Phys. Rev. Lett. \textbf{89}, 187903 (2002).
\bibitem{LYX2003} You, L., Yi, X. X., Su, X. H.: Quantum logic between atoms inside a high-Q optical cavity. Phys. Rev. A \textbf{67}, 032308 (2003).
\bibitem{KI2001} Ichimura, K.: A simple frequency-domain quantum computer with ions in a crystal coupled to a cavity mode. Opt. Commun. \textbf{196}, 119 (2001).
\bibitem{MSS2002} Shahriar, M. S., Hemmer, P. R., Lloyd, S., Bhatia, P. S., Craig, A. E.: Solid-state quantum computing using spectral holes. Phys. Rev. A \textbf{66}, 032301 (2002).
\bibitem{SMS2005} Spillane, S. M., Kippenberg, T. J., Vahala, K. J., Goh, K. W.,
Wilcut, E., Kimble, H. J.: Ultrahigh-Q toroidal microresonators
for cavity quantum electrodynamics. Phys. Rev. A \textbf{71},
013817 (2005).
\end{thebibliography}
\end{document}